\documentclass[twocolumn,pra,twocolumn,superscriptaddress]{revtex4}
\usepackage[latin9]{inputenc}
\usepackage{color}
\usepackage{amsmath}
\usepackage{amssymb}
\usepackage{graphicx}
\usepackage{tikz}
\usepackage{mathrsfs} 
\usepackage{float}
\usepackage{bbm}
\usepackage{epstopdf}
\usepackage{epsfig}
\usepackage{verbatim}
\usepackage{array}
\usepackage{setspace}
\usepackage{times}
\usepackage{multirow}

\usepackage[unicode=true,
bookmarks=true,bookmarksnumbered=false,bookmarksopen=false,
breaklinks=false,pdfborder={0 0 1},backref=false,colorlinks=true]
{hyperref}
\hypersetup{
	linkcolor=magenta, urlcolor=blue, citecolor=blue, pdfstartview={FitH}, hyperfootnotes=false, unicode=true}

\newcolumntype{C}[1]{>{\centering\arraybackslash$}p{#1}<{$}}

\newcommand{\bra}[1]{\langle#1|}
\newcommand{\ket}[1]{|#1\rangle}

\begin{document}

\title{Geometric two-qubit gates in silicon-based double quantum dots}

\author{Yong-Yang Lu}

\affiliation{School of Physical Science and Technology, Guangxi University, Nanning 530004, China}
\affiliation{Guangxi Key Laboratory for Relativistic Astrophysics,
	School of Physical Science and Technology, Guangxi University, Nanning 530004, China}



\author{Kejin Wei}

\author{Chengxian Zhang}
\email[Electronic address:]{cxzhang@gxu.edu.cn}
\affiliation{School of Physical Science and Technology, Guangxi University, Nanning 530004, China}
\affiliation{Guangxi Key Laboratory for Relativistic Astrophysics,
	School of Physical Science and Technology, Guangxi University, Nanning 530004, China}

\date{\today}

\begin{abstract}

Achieving high-fidelity two-qubit gates is crucial for spin qubits in silicon double quantum dots.  However, the two-qubit gates in experiments are easily suffered from charge noise, which is still a key challenge. Geometric gates which implement gate operations employing pure geometric phase are believed to be a powerful way to realize robust control. In this work, we theoretically propose feasible strategy to implement geometric two-qubit gates for silicon-based spin qubits considering experimental control environments. By working in the suitable region where the local magnetic field gradient  is much larger than the exchange interaction, we are able to implement entangling and non-entangling geometric gates via analytical and numerical methods.  It is found that the implemented geometric gates can obtain  fidelities surpassing 99\% for the noise level related to the experiments. Also, they can outperform the dynamical opertations. Our work paves a way to implement high-fidelity geometric gate for spin qubits in silicon.

\end{abstract}

\maketitle

\section{Introduction}
Achieving high-fidelity gate operations is pivotal for the realization of fault-tolerant quantum computing and various quantum tasks.  In the pursuit of viable quantum computer hardware, spin qubits encoded in silicon double quantum dots (DQDs) attract many attentions due to its potential scalability, all electrical control and combination with the modern semiconductor technology \cite{LossPRA_1998,BurkardRMP_2023} . Over the past years, significant progress has been made, showing single-qubit gates with fidelities exceeding 99.9\% \cite{YonedaNaNan_2018,YangNaEle_2019,CerfontaineNaCom_2020,XueNature_2021}. Meanwhile, several groups have realized two-qubit gate operations with fidelities above 99\% \cite{Noiri.22,Madzik.22,Xue.22}.  Despite these advancements, silicon-based spin qubits continue to be affected by various noises. The typical two noise sources are the charge noise \cite{Yoneda.18} and the nuclear noise \cite{Huang.19}. In experiments, the nuclear noise has been effectively reduced by using the isotopic purification method \cite{Huang.19}. On the other hand, the charge noise will lead to fluctuations in exchange interaction and resonance frequencies \cite{YonedaNaNan_2018,ChanPRApp_2018}, where both single- and two-qubit gates can then be affected.  The charge noise in experiments still cannot be neglected safely, especially for the two-qubit case. Nowadays, approaches that can suppress noise on qubit operations and mitigate the gate errors are still required.

Recent years, many methods have been proposed to mitigate gate errors, such as composite quantum gates \cite{WangNatureCom_2012,BandoJPSJ_2013,ZhangPRL_2017,GevorgyanPRA_2021}, time-optimal quantum gates \cite{WangPRL_2015,GengPRL_2016,DridiPRL_2020}, and geometric gates \cite{PachosPRA_1999,ZanardiPLA_1999,DuanScience_2001,ZhuPRL_2002,ZhuPRL_2003,ZhaoPRA_2017}. Among these methods, Geometric gates  stand out  for achieving high-fidelity quantum operations.  Different from other traditional dynamical methods implemented by using the dynamical phase, the construction of geometric gates typically requires two fundamental criteria \cite{SjoqvistIJQC_2015}. One is to cancel the dynamical phase and the other is to realize cyclic evolution in the parameter space.  Since the geometric gates implement gate operations using only pure geometric phase, it might be able to effectively mitigate fluctuations in the control Hamiltonian due to the intrinsic global property \cite{ZHANG.23}.

Geometric gate can be constructed using either the Berry phase \cite{Berry_1984} or the Aharonov-Anandan (AA) phase \cite{AharonovPRL_1987} (see Appendix.~\ref{append}), corresponding to adiabatic and nonadiabatic evolution, respectively. In early proposals, many techniques employed the Berry phase to implement gate operations. However, the adiabatic condition associated with the Berry phase necessitates excessively long evolution times, leading to heightened decoherence in the gates. In contrast, geometric gates based on the AA phase enable shorter evolution time, facilitating faster experimental implementation. AA-based universal geometric quantum gates have been successfully demonstrated in various platforms, such as superconducting circuits \cite{AbdumalikovNa_2013,ChenPRApp_2018,XuPRL_2018,LiuPRL_2019,EggerPRApp_2019,XuPRL_2020,ChenPRApp_2020}, Nitrogen-vacancy centers in diamond \cite{SekiguchiNa_2017,ZhouNaturePhy_2017}, trapped ions \cite{AiPRApp_2020,AiFR_2022}.  In Ref.~ \cite{Ma.24}, single-qubit geometric gate operations are experimentally demonstrated possible in silicon DQDs. Nevertheless, experimental realization of two-qubit geometric gate operations  are still scarce in literature.  Part of the reason is that  an effective two-level system with controllable Rabi frequency and phase of the microwave field is crucial to achieve pure geometric phase. It is easy for the single-qubit case to meet this condition. But, for the two-qubit case, it requires block diagonalization of the full Hamiltonian. Meanwhile, the other cross terms owing to the microwave field can be neglected with good approximation so as to obtain the effective two-level system. Another reason is that it is hard to design precise evolution for different subspaces, both of which satisfy desired evolution for geometric operations.

In this study, we propose feasible strategy to implement geometric two-qubit gates for silicon-based spin qubits considering experimental operating environments. We first systematically study the two-qubit Hamiltonian in silicon DQDs, aiming to realizing good block diagonalization. We then subtly design how to  implement geometric gates with real experimental limitations. By working in the suitable regime, i.e.,  the exchange interaction is far less than the local magnetic field gradient, we are able to implement geometric two-qubit gates including entangling and non-entangling operations.  On one hand, by driving the oscillating magnetic field using EDSR with square pulse, we can realize entangling CZ gate and non-entangling $\sqrt{\rm{CNOT}}$ gate via analytical and numerical methods, respectively. On the other hand, by using the microwave-driven signal on the exchange interaction with pulse shape that is friendly to the experiments, we can implement both iSWAP gate and SWAP gate. We also perform simulation to analyze the performance of the  geometric two-qubit gates.  Simulation result indicates the achieved geometric operations can achieve  fidelities surpassing 99\% for the noise level related to the experiments. We also find that the geometric gate can outperform the dynamical operations.

\section{Model}
\label{sec2}
\begin{figure}[htbp]
    \centering
\includegraphics[width=8.5cm,height=5cm]{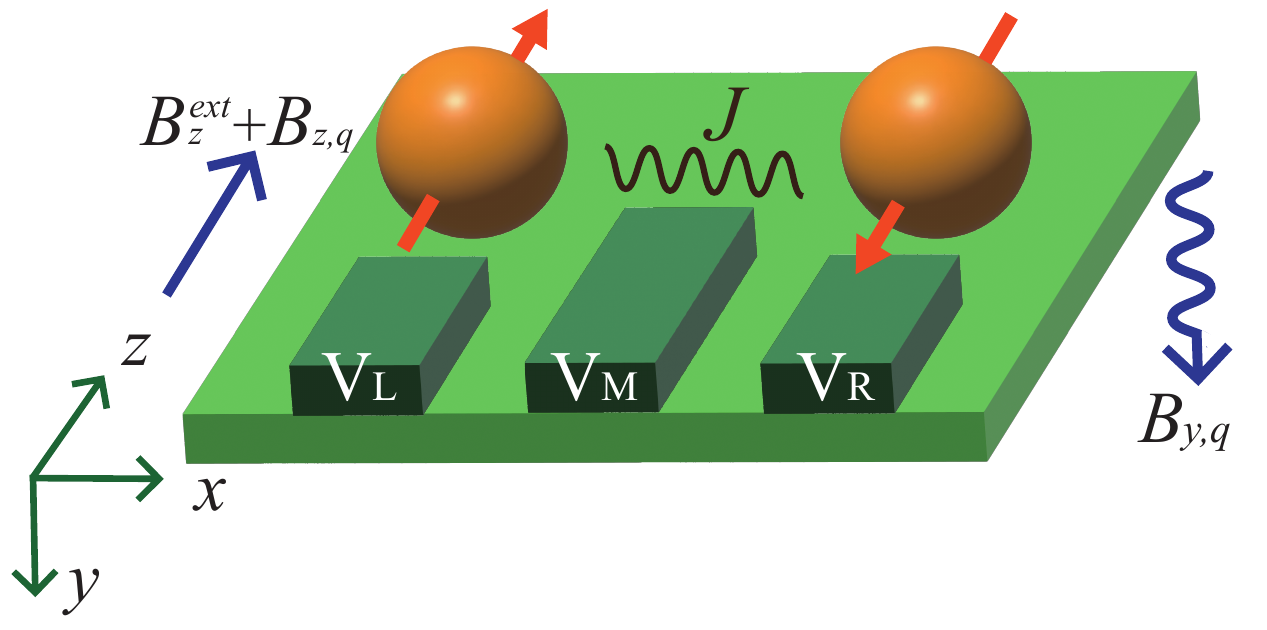}
    \caption{ Spin qubits in the silicon-based DQDs.  Two-qubit gates can be implemented by manipulating the oscillating
    	magnetic field $\mathbf{B}$ induced by the microwave and the exchange interaction $J$ via the gate voltages $V_L$ ($V_R$) or $V_M$.}
    \label{Fig1}
    \end{figure} 

As shown in Fig.~\ref{Fig1}, the system under investigation \cite{Zajac.18, Russ.18} comprises two electrons  which are operated in the (1,1) regime in the silicon DQDs.  Here, $(n_{L}, n_{R})$ denotes the number of the electron confined in the left and right dot, respectively. The system can be described using the Heisenberg Hamiltonian \cite{VargasPRB_2019}  
\begin{flalign}
	\label{eq1}
	H(t)=J(t)\left(\mathbf{S}_L \cdot \mathbf{S}_R-1 / 4\right)+\mathbf{S}_L \cdot \mathbf{B}_L+\mathbf{S}_R \cdot \mathbf{B}_R.
\end{flalign}
Here, $J(t)$ is the Heisenberg exchange interaction, which signifies the interacting strength between the left spin $\mathbf{S}_L$ in the left dot and right spin  $\mathbf{S}_R$ in the right dot. Experimentally, it can be controlled easily by biasing the detuning value $\epsilon$ via the gate voltages ($V_{L}$ and $V_{R}$) or tunning the tunneling element $t_{m}$ via barrier gate voltage $V_{M}$. $\mathbf{B}_L=(0, B_{y,L}(t), B_{z,L}(t)+B^{\rm{ext}}_{z})^T$ and $\mathbf{B}_R=(0, B_{y,R}(t), B_{z,R}(t)+B^{\rm{ext}}_{z})^T$ denote the magnetic fields (in energy units) acting on the spins in the left and right dots, respectively. $B_{y,q}(t)=B_{y,q}^{0}+B_{y,q}^{1}\cos(\omega t +\varphi)$ ($q=L,R$) represents the transverse AC driving oscillating field with a static part $B_{y,q}^{0}$ and a variable amplitude $B_{y,q}^{1}$. In addition, $\omega$ and $\varphi$ are the frequency and phase of the oscillating field. In the experiments, the oscillating field can be generated either by using the technologies of electron spin resonance (ESR) \cite{Huang.19} or electric-dipole spin resonance (EDSR) \cite{Zajac.18}.  In this work, we consider the later. $B^{\rm{ext}}_z$ signifies the homogeneous external magnetic field in the $z$ direction, which can lift the spin degeneracy. $B_{z,q}(t)=B_{z,q}^{0}+B_{z,q}^{1}(t)$ denotes the local magnetic field, which is time-dependent caused by the change in the electron positions  when the energy barrier between dots is altered by the middle metal gate \cite{Zajac.18,VargasPRB_2019}.  In this way, each spin can be individually addressed with different resonant frequency to perform single-qubit gates.  In experiments, the Zeeman splitting of the two qubits can be modulated in three particular regimes. In the regime where the exchange interaction is much larger than the local magnetic
field gradient, namely, $J \gg\left(B_{z, R}-B_{z, L}\right)$, the eigenstates of Eq.~(\ref{eq1}) are approximately the singlet state $(|\uparrow \downarrow\rangle-|\downarrow \uparrow\rangle)/\sqrt{2}$ and the triplet states  $\{|\uparrow \uparrow\rangle, (|\uparrow \downarrow\rangle+|\downarrow \uparrow\rangle)/\sqrt{2}, |\downarrow \downarrow\rangle\}$, which is beneficial to implement a two-qubit $\sqrt{\rm{\small\textsc{SWAP}}}$ gate. Further, a $\rm{\small\textsc{CNOT }}$ gate can be achieved by combining two $\sqrt{\rm{\small\textsc{SWAP}}}$ gates and several single-qubit gates \cite{LossPRA_1998}. While when the exchange interaction is negligible, i.e., $J\sim0$, single-qubit operations can be realized by matching  $\omega$ with the resonance frequency for each dot.  Particularly, by adjusting the phase $\varphi$ allows precise control over the single-qubit rotation axis within the equatorial plane of the Bloch sphere.  In this work,  our analysis focuses on the regime where the Zeeman splitting of the qubits significantly exceeds the strength of the exchange interaction, i.e., $J \ll\left(B_{z, R}-B_{z, L}\right)$. In this regime, the two-qubit eigenstates are the product states of the two spins, as shown below in Eq.~(\ref{eq5}). In fact, it is typical for DQDs systems to implement entangling gates in the presence of a micromagnet.

\section{CZ Gate by modulating magnetic field}
\label{sec3}
\subsection{Effective Hamiltonian}
In the normal basis $\{\ket{\uparrow \uparrow},|\downarrow \uparrow\rangle,|\uparrow \downarrow\rangle,\ket{\downarrow \downarrow}\}$, the Hamiltonian in Eq.~(\ref{eq1}) is written as
\begin{widetext}
\begin{equation}
{H'} =  \begin{pmatrix}
E_z + E^1_{z} & -\frac{i B_{y,L}}{2} & -\frac{i B_{y,R}}{2} & 0 \\
\frac{i B_{y,L}}{2} & \frac{1}{2}(-J + \Delta E_z + \Delta E^1_{z}) & \frac{J}{2} & -\frac{i B_{y,R}}{2} \\
\frac{i B_{y,R}}{2} & \frac{J}{2} & \frac{1}{2}(-J - \Delta E_z - \Delta E^1_{z}) & -\frac{i B_{y,L}}{2} \\
0 & \frac{i B_{y,R}}{2} & \frac{i B_{y,L}}{2} & -E_z - E^1_{z}
\end{pmatrix}. 
\label{eq2}
\end{equation}
\end{widetext}
Here, $E_z = B^{\rm{ext}}_z + (B_{z,L}^{0} + B_{z,R}^{0} )/2$ represents the average Zeeman splitting, and $\Delta E_z = B_{z,R}^{0}-B_{z,L}^{0}$ indicates the Zeeman splitting between the two dots. $E^1_z = (B_{z,R}^{1} + B_{z,L}^{1})/2$ and $\Delta E^1_{z}= B_{z,R}^{1}-B_{z,L}^{1}$ denote the Zeeman shifts.  It is noted that $\Delta E_z+\Delta E^1_{z}= B_{z, R}-B_{z, L}$ actually reflects the magnetic field gradient of the two dots, such that we also have $J \ll\left(\Delta E_z+\Delta E^1_{z}\right)$.


To facilitate the analytical derivation of geometric entangling operations, we move to the rotating frame using the transformation $H_{\mathrm{rot}}(t)=U_\omega H' U_\omega^{\dagger}-i U_\omega \dot{U}_\omega^{\dagger}$ with $U_\omega=\exp \left[i \omega t\left(S_{z, L}+S_{z, R}\right) / \hbar\right]$. Considering rotating wave approximation (RWA), the Hamiltonian is further expressed as
\begin{widetext}
\begin{equation}
\label{eqrot}
H_{\mathrm{rot}} ^{1}=
\begin{pmatrix}
E_z + E^1_z - \omega & -\frac{iB_{y,L}^{1}e^{-i\varphi}}{4} &- \frac{iB_{y,R}^{1}e^{i\varphi}}{4} & 0 \\
\frac{iB_{y,L}^{1}e^{i\varphi}}{4} & -\frac{1}{2} \left( -J + \Delta E_z + \Delta E^1_z \right) & \frac{J}{2} & -\frac{iB_{y,R}^{1}e^{-i\varphi}}{4} \\
\frac{iB_{y,R}^{1}e^{i\varphi}}{4} & \frac{J}{2} & \frac{1}{2} \left( -J - \Delta E_z - \Delta E^1_z \right) & -\frac{iB_{y,L}^{1}e^{-i\varphi}}{4} \\
0 & \frac{iB_{y,R}^{1}e^{i\varphi}}{4} & \frac{iB_{y,L}^{1}e^{i\varphi}}{4} & -E_z - E^1_z + \omega
\end{pmatrix}.
\end{equation}
\end{widetext}
Using the Hamiltonian's instantaneous adiabatic eigenstates \\ $\{\ket{\uparrow \uparrow},\widetilde{|\downarrow \uparrow\rangle},|\widetilde{\uparrow \downarrow}\rangle,|\downarrow \downarrow\rangle\}$  as the new basis, the Hamiltonian can be further rewritten as
\begin{widetext}
	\begin{equation}
		\label{eq4}
		\begin{aligned}
			H_{\mathrm{rot}} ^{2}=
			\begin{pmatrix}
				E_z + E^1_z - \omega & -\frac{iB_{y,L}^{1}e^{-i\varphi}}{4} & 
				-\frac{iB_{y,R}^{1}e^{-i\varphi}}{4} & 0 \\
				\frac{iB_{y,L}^{1}e^{i\varphi}}{4} & \frac{1}{2} \left( -J + \sqrt{J^2 + (\Delta E_z + \Delta E^1_z)^2}\right)
				& 0 & -\frac{iB_{y,R}^{1}e^{-i\varphi}}{4} \\
				\frac{iB_{y,R}^{1}e^{i\varphi}}{4} & 0 & \frac{1}{2} \left( -J - \sqrt{J^2 + (\Delta E_z + \Delta E^1_z)^2}\right) & -\frac{iB_{y,L}^{1}e^{-i\varphi}}{4} \\
				0 & \frac{iB_{y,R}^{1}e^{i\varphi}}{4} & \frac{iB_{y,L}^{1}e^{i\varphi}}{4} & -E_z - E^1_z + \omega
			\end{pmatrix}.
		\end{aligned}
	\end{equation}
\end{widetext}
Here, 
\begin{equation}
\label{eq5}
\begin{aligned}
\widetilde{|\downarrow \uparrow\rangle}&=\sin\theta \ket{\downarrow\uparrow}   +\cos\theta \ket{\uparrow\downarrow},\\
 \ket{\widetilde{\uparrow\downarrow}}&=-\cos\theta \ket{\downarrow\uparrow} + \sin\theta \ket{\uparrow\downarrow},
\end{aligned}
\end{equation}
where $\sin \theta = \frac{\alpha+\beta}{\sqrt{(\alpha+\beta)^2+1}}$ and $\cos\theta = \frac{1}{\sqrt{(\alpha+\beta)^2+1}}$ with $\alpha=\frac{\Delta E_z + \Delta E^1_z}{J}$ and $ \beta=\frac{\sqrt{J^2+(\Delta E_z + \Delta E^1_z)^2}}{J}$.  As stated above, since we consider $J \ll\left(B_{z, R}-B_{z, L}\right)$ in this work,  we can simplify the Hamiltonian using the approximation, namely, $\cos\theta\sim0$, $\sin\theta\sim1$ and $\sqrt{J^2+(\Delta E_z + \Delta E^1_z)^2}\sim\Delta E_z + \Delta E^1_z+\frac{J^2}{2(\Delta E_z + \Delta E^1_z)}$.   In addition, when designing the entangling gate, the terms related to $B_{y,R}^{1}$ can be safely removed. While its  effect on other gates are shown below in Fig.~\ref{Fig5}. Note that although we condider the approximation, we actually use the full Hamiltonian when numerically calculating the fidelity of the gates as shown below. By taking the resonance frequency as $\omega=E_z+E^1_z+\frac{1}{2}\left( J - \Delta E_z - \Delta E^1_z - \frac{J^2}{2(\Delta E_z+\Delta E^1_z)}\right)$, the Hamiltonian in Eq.~(\ref{eq4}) can be further written in a diagonalized form
\begin{equation}
	\begin{aligned}
		\label{eq6}
		H_{\mathrm{rot}}^{3} &=C_{1}\tilde{I}-C_{2}\tilde{i}+
		\begin{pmatrix}
			0 & \frac{h}{2}e^{-i\phi} & 0 & 0 \\
			\frac{h}{2}e^{i\phi} & 0
			& 0 & 0 \\
			0 & 0 & -\frac{J}{2}   & \frac{h}{2}e^{-i\phi} \\
			0 & 0 & \frac{h}{2}e^{i\phi} & \frac{J}{2}
		\end{pmatrix}.\\
	\end{aligned}
\end{equation}
Here, $\tilde{I}$ is the identity operator under the instantaneous adiabatic eigenstates basis, $\tilde{i}=|\widetilde{\uparrow \downarrow}\rangle   \langle \widetilde{\uparrow \downarrow|}+|\downarrow \downarrow\rangle   \langle\downarrow \downarrow|$, $C_{1}=\frac{1}{2}\left( \Delta E_z +\Delta E^1_z  \right)+ \frac{J^{2}}{2(\Delta E_z+\Delta E^1_z)}-J$ and $C_{2}=\frac{1}{2}\left( \Delta E_z +\Delta E^1_z  \right)+ \frac{J^{2}}{2(\Delta E_z+\Delta E^1_z)}-\frac{J}{2}$. In addition, we set $h=B_{y,L}^{1}/2$ and $\phi=\varphi+\pi/2$ for simplicity. It is clear that $C_{1}\tilde{I}$ can induce a global phase factor, while $C_{2}\tilde{i}$ can also accumulate a phase factor within the subspace spanned by $\{|\widetilde{\uparrow \downarrow}\rangle,     |\downarrow \downarrow\rangle  \}$. When neglecting the global phase factor, the Hamiltonian can be expressed as the product of two subspaces as $H_{\mathrm{rot}}^{4} =H_{S_1}\oplus (-C_{2}\tilde{i}+H_{S_2})$, where 
\begin{equation}
	\begin{aligned}
		\label{eq6}
		H_{S_1}&=
		\begin{pmatrix}
			0 & \frac{h}{2}e^{-i\phi}  \\
			\frac{h}{2}e^{i\phi}  &   0
		\end{pmatrix},
	\end{aligned}
\end{equation}
and
\begin{equation}
	\begin{aligned}
		\label{eq6}
		H_{S_2} &=
		\begin{pmatrix}
		-\frac{J}{2}   & \frac{h}{2}e^{-i\phi} \\
		\frac{h}{2}e^{i\phi} & \frac{J}{2}
		\end{pmatrix}.
	\end{aligned}
\end{equation} Here, $S_1$ and $S_2$ represent the subspaces spanned by $\{\ket{\uparrow \uparrow},\widetilde{|\downarrow \uparrow\rangle}\}$ and $\{|\widetilde{\uparrow \downarrow}\rangle,|\downarrow \downarrow\rangle\}$, respectively.  Then, we can design a geometric  identity operator in the subspace $S_1$ and a $Z(\pi)$ rotation in $S_2$, such that a  Controlled-Z ($\rm{CZ}$) gate can be realized, the deatil of which is as shown below in section.~\ref{GQC}.


\begin{figure}
	\flushleft
	\includegraphics[width=8cm,height=4.5cm]{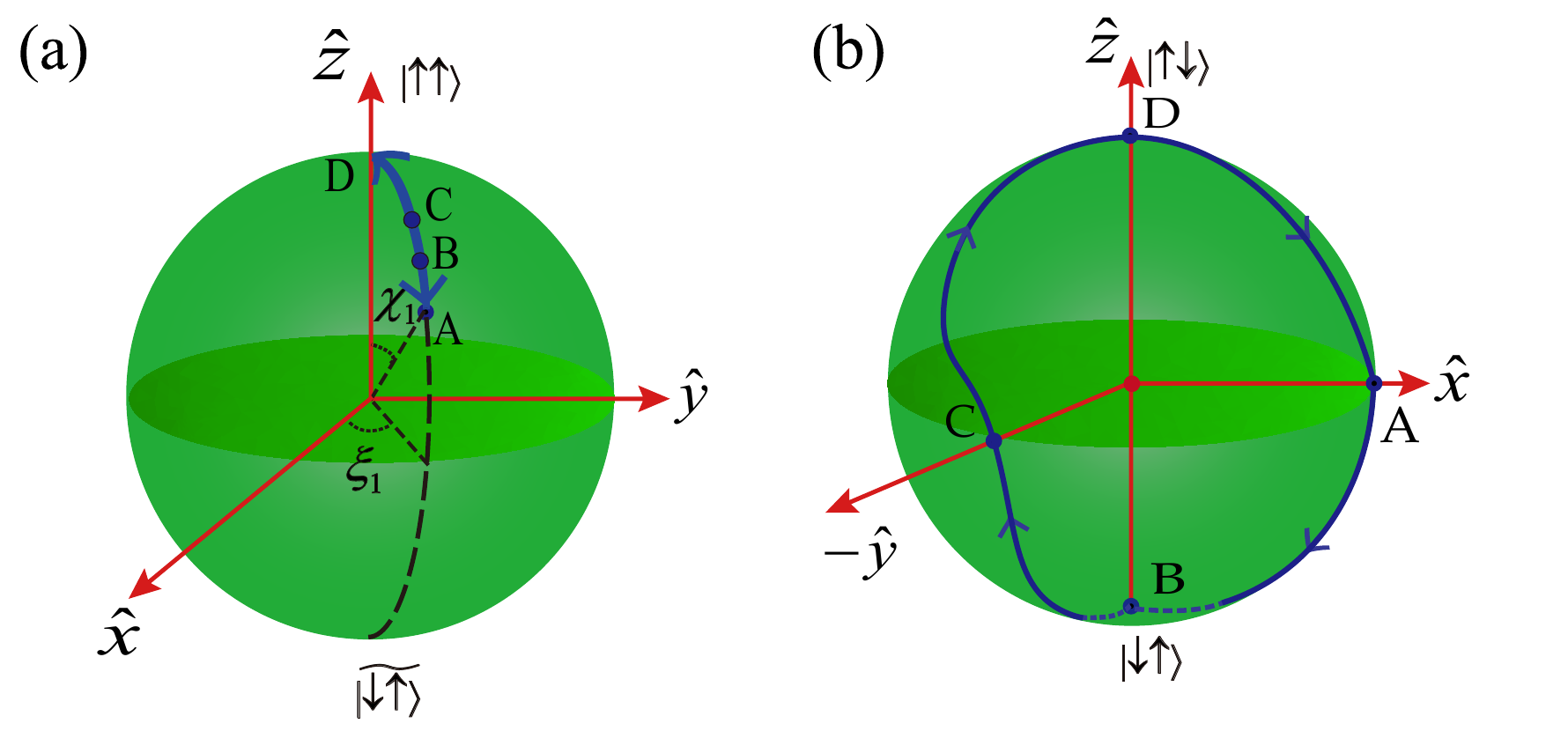}
	\caption{The evolution paths of the dressed states for the geometric gates on the Bloch sphere. (a) denotes evolution of a geometric identity operation for the CZ gate in the subspace $S_1$, which corresponds to Eq.~(\ref{path1}). (b) denotes the evolution of a NOT gate for the iSWAP gate corresponding to Eq.~(\ref{fequas}).}
	\label{Fig2}
\end{figure} 

\subsection{Geometric $\rm{CZ}$ gate}
\label{GQC}
We begin by introducing how to implement geometric gates \cite{JiAQT_2021,ChenPRApp_2022,MaPRApp_2023} based on the reduced Hamiltonian $H_{S_1}$ in the subspace $S_1$. According to Lewis-Riesenfeld invariant methods\cite{LewisJMP_2003,ChenPRA_2011,RuschhauptNJP_2012}, a set of orthogonal dressed states in $S_1$ are chosen as 
\begin{equation}
	\begin{aligned}
		\label{eq7}
\ket{\psi_{+}(t)}&=e^{if_{+}(t)}(\cos[\chi(t)/2]\ket{\uparrow\uparrow}+\sin[\chi(t)/2]e^{i\xi(t)}\ket{\widetilde{\downarrow\uparrow}}),\\ \ket{\psi_{-}(t)}&=e^{if_{-}(t)}(\sin[\chi(t)/2]e^{-i\xi(t)}\ket{\widetilde{\uparrow\uparrow}}-\cos[\chi(t)/2]\ket{\widetilde{\downarrow\uparrow}}),
	\end{aligned}
\end{equation}
where $f_i(t)\,(i=+,-)$  is the global phase factor, while $\xi(t)$ and $\chi(t)$ represent the polar and azimuthal angles of state vector on the Bloch sphere. By inserting $\ket{\psi_{i}(t)}$ into the Schr\"odinger equation governed by $H_{S_1}$, $\chi(t)$ and $\xi(t)$ can be determined as
\begin{equation}
    \begin{aligned}
    		\label{eq8}
        \dot{\chi}(t) &= h\sin(\phi-\xi),\\
        \dot{\xi}(t)&=-h \cot\chi\cos(\phi-\xi).
    \end{aligned}
\end{equation}
Thus, the parameters $\{h, \phi \}$ in $H_{S_1}$ can be inversely determined as
\begin{equation}
\begin{aligned}
		\label{eq9}
 h(t)&= \pm \sqrt{\dot{\phi}^2+\dot{\chi} \cot ^2 \chi} \tan \chi, \\
\phi(t)&=\chi-\arctan \left( \pm \frac{\dot{\chi} \cot \chi}{2 \dot{\xi}}\right).
\end{aligned}
\end{equation} Therefore, we can design the required Hamiltonian $H_{S_1}$ for any parameter set $\{\chi(t), \xi(t) \}$, and this is beneficial for designing the geometric gate as shown below. 

On the other hand,  after a period $\tau$, the dressed state $\ket{\psi_{+}(t)}$ ($\ket{\psi_{-}(t)}$)  fullfills a cyclic evolution. Accordingly, it obtains a global phase factor as $f_{+}(\tau)$ ($f_{-}(\tau)$) with $f_{+}(\tau)=-f_{-}(\tau)=\gamma=-\frac{1}{2} \int \dot{\xi}(t) d t+\frac{1}{2} \int \frac{\dot{\xi}(t)}{\cos \chi(t)} d t$.  $\gamma$ can be seperated into two parts, namely, the dynamical part 
\begin{equation}
    \begin{aligned}
    		\label{eq10}
        \gamma_d&=-\int_0^\tau\bra{\psi_{+}(t)}H_{S_1}\ket{\psi_{+}(t)}dt\\
        &=\int_0^\tau \frac{\sin^2\chi(t)\dot{\xi}(t)}{\cos\chi(t)}dt, 
    \end{aligned}
\end{equation}
and the geometric part as $\gamma_{g}=\gamma-\gamma_{d}$.  In addition,  the obtained operator at the final time in the subspace $S_{1}$ can be expressed by the dressed states as
\begin{widetext}
\begin{equation}
	\begin{aligned}
		U(\tau)&=\left|\psi_{+}(\tau)\right\rangle\left\langle\psi_{+}(0)|+| \psi_{-}(\tau)\right\rangle\left\langle\psi_{-}(0)\right| \\
		& =\left(\begin{array}{cc}
			e^{-i \xi_{-}}(\cos \gamma^{\prime} \cos \chi_{-}+i \sin \gamma^{\prime} \cos \chi_{+}) & e^{-i \xi_{+}}(-\cos \gamma^{\prime} \sin \chi_{-}+i \sin \gamma^{\prime} \sin \chi_{+}) \\
			e^{i \xi_{+}}(\cos \gamma^{\prime} \sin \chi_{-}+i \sin \gamma^{\prime} \sin \chi_{+})& e^{i \xi_{-}}(\cos \gamma^{\prime} \cos \chi_{-}-i \sin \gamma^{\prime} \cos \chi_{+})
		\end{array}\right),
	\end{aligned}
\end{equation}
\end{widetext}
where  $\xi_{ \pm}=\frac{\xi(\tau) \pm \xi(0)}{2}$, $\chi_{ \pm}=\frac{\chi(\tau) \pm \chi(0)}{2}$,  and $\gamma^\prime=\gamma_g+\xi_{-}$.

\begin{figure}
	\centering
	\includegraphics[width=0.96\columnwidth]{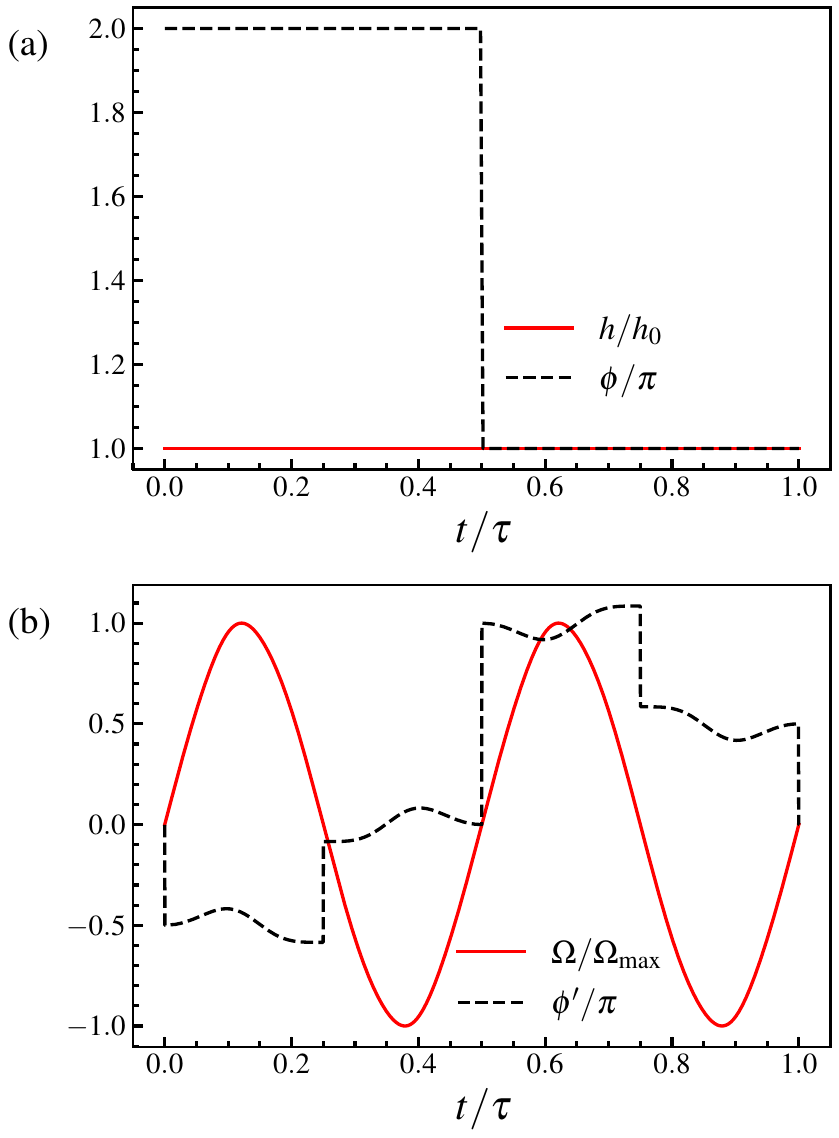}
	\caption{(a) The pulse shapes of driving fields, (a) $h(t)$ and $\varphi(t)$ for the CZ gate and (b) $\Omega(t)$ and $\phi'(t)$ for the iSWAP gate.}
	\label{Fig3}
\end{figure}

The key to constructing a geometric gate is to cancel the dynamical phase $\gamma_{d}$. According to Eq.~(\ref{eq10}), this can be realized by taking $\dot{\xi}(t)=0$ for a given evolution duration. This in turn implies that the dressed state evolves always along the longitude line, except for the singular points (the north and sole poles of the Bloch sphere) \cite{Fang.24}.  To construct the CZ gate, it requries that $U(\tau)$ in the subspace $S_{1}$  is equivalent to an effective identity operater at final time $\tau$. To this end, we take the parameters as $\chi(0)=\chi(\tau)=\chi_{1}=\pi/25$,  and $\xi(0)=\xi(\tau)=\xi_{1}=3\pi/2$.  The evolution path of the dressed state $\ket{\psi_{+}(t)}$ can be visualized on the Bloch sphere as shown in Fig.~\ref{Fig2}(a) as $A \rightarrow D \rightarrow A$. Specifically, at the beginning, it starts from point A with coordinates as $(\chi_1,\xi_1)$. Then, it travels to the north pole B $(0,\xi_1)$ at time $\tau_{1}$ along the longitude line. Then, it goes back to point A along the same path. According to Eq.~(\ref{eq9}), the parameters $h(t)$ and $\phi(t)$ in $H_{S_1}$ for these two evolution segments satisfy with
\begin{equation}
	\label{path1}
	\begin{aligned}
		\int_{0}^{\tau_1}h(t)dt&=\frac{\chi_1}{2},\quad \phi(t)=\phi_1=\xi_1+\frac{\pi}{2},\\
		\int_{\tau_1}^{\tau}h(t)dt&=\frac{\chi_1}{2},\quad \phi(t)=\phi_2=\xi_1-\frac{\pi}{2}.
	\end{aligned}
\end{equation}
The shapes of $h(t)$ and $\phi(t)$ with respect to time are shown in Fig.~\ref{Fig3}(a). Because $h$ and $J$ related to $H_{S_2} $ are not commutative with each other, the pulse shape of $h(t)$ are better to chose as square form, i.e., $h(t)=h_{0}$, which is helpful to design precise evolution in subspace $S_2$.  Meanwhile, $\phi(t)$ is also with two distinct values. In this way, the resulted gate $U(\tau,0)=U_{2}(\tau,\tau_1)U_{1}(\tau_1,0)=[1, 1]^{\rm{T}}$ is thus an identity operator in $S_1$.  On the other hand, during the evolution process we have $\dot{\xi}(t)=0$, it is clear that $U(\tau,0)$ is pure geometric.  Then, the remaining question is how to solve the value of the exchange interaction in the subspace $S_{2}$. 
A simple way is to set it to be constant during the evolution. Thus it can be determined as $J=37.4879\ h_{0}$ via numerically solving $R(J,\phi_2, \tau_{2})R(J,\phi_1, \tau_{1})=Z(\pi)=[-i, i]^{\rm{T}}$  within subspace $S_2$, where $R(J,\phi_i, \tau_{i})=e^{-\frac{i}{\hbar}\int H_{S_2}(t) dt}$ ($i$=1,2) denotes the evolution operator in $S_{2}$.  Meanwhile, it requires that $C_{2}\times(\tau_{1}+\tau_{2})=\frac{n\pi}{2}$ with $n$ being an odd number.  Then in the rotating frame and in the basis  $\{\ket{\uparrow \uparrow},\widetilde{|\downarrow \uparrow\rangle},|\widetilde{\uparrow \downarrow}\rangle,|\downarrow \downarrow\rangle\}$ we obtain the two-qubit gate as 
\begin{equation}
	\begin{aligned}
		U_{\rm{ad}}&=i\begin{pmatrix}
			-1& 0 & 0 & 0 \\
			0 &-1 & 0 & 0 \\
			0 & 0& -1 &0 \\
			0 & 0 & 0 & 1
		\end{pmatrix}.
 \end{aligned}
\end{equation}
Finally, we have a CZ gate when we are back to the normal basis $\{\ket{\uparrow \uparrow},\widetilde{|\uparrow \downarrow\rangle},|\widetilde{\downarrow \uparrow}\rangle,|\downarrow \downarrow\rangle\}$
\begin{equation}
	\begin{aligned}
		U_{\rm{cz}}&=-i\begin{pmatrix}
		1 & 0 & 0 & 0 \\
			0 & 1 & 0 & 0 \\
			0 & 0& 1 &0 \\
			0 & 0 & 0 & -1
		\end{pmatrix},
	\end{aligned}
\end{equation}
up to a global phase factor determined by $C_1$ and the total evolution time $\tau$.

\begin{figure}
	\centering
	\includegraphics[width=0.96\columnwidth]{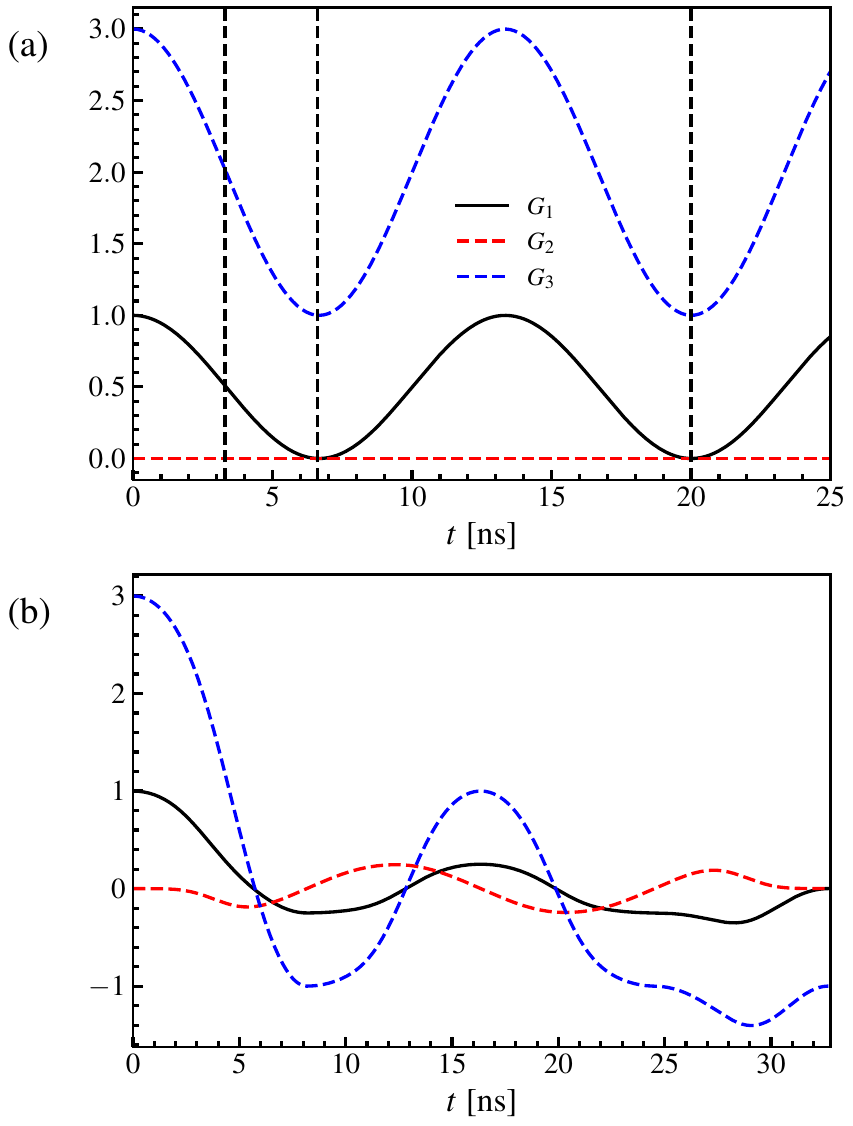}
	\caption{Local invariants evolves with respect to time, where (a) is for CZ gate and (b) for iSWAP gate.}
	\label{Fig4}
\end{figure}

\begin{figure*}
	\centering
	\includegraphics[width=1.98\columnwidth]{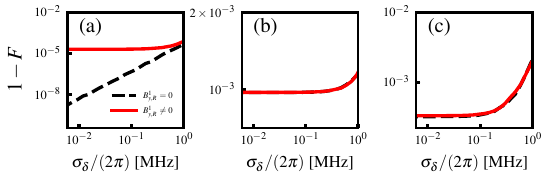}
	\caption{The infidelity for the implemented geometric operations, where (a) is for the $\sqrt{\rm{CNOT}}$ gate, (b) and (c) for the two CZ gates, i.e., $U'_{\rm{cz}}$ and $U_{\rm{cz}}$.}
	\label{Fig5}
\end{figure*}

\begin{table}
\begin{tabular}{lccc}
	\hline \hline Gate & $G_1$ & $G_2$ & $G_3$ \\
	\hline $\sqrt{\rm{CNOT}}$ & 0.5 & 0 & 2 \\
	CZ (CNOT) & 0 & 0 & 1 \\
	iSWAP & 0 & 0 & -1 \\
	SWAP & -1 & 0 & -3 \\
	\hline \hline
\end{tabular}
\caption{Local invariants for the implemented geometric two-qubit gates in this work.}
\label{ta1}
\end{table}

In fact, during the whole evolution we can further numerically determine  what types of entangling gates can be generated. To this end, we calculate the so-called  ``local invariants'' which are represented by  a set of three real numbers \cite{MakhlinQIP_2002,ZhangPRA_2003}. The collection of local invariants can be expressed as follows: 
\begin{equation}
	\begin{aligned}
		G_1 & =\operatorname{Re}\left[\frac{\operatorname{tr}^2[M(U)]}{16 \operatorname{det} U}\right], \\
		G_2 & =\operatorname{Im}\left[\frac{\operatorname{tr}^2[M(U)]}{16 \operatorname{det} U}\right], \\
		G_3 & =\frac{\operatorname{tr}^2[M(U)]-\operatorname{tr}\left[M^2(U)\right]}{4 \operatorname{det} U} .
	\end{aligned}
\end{equation}
These invariants  can reveal the non-local or entangling property of the considered two-qubit operations.  Table.~\ref{ta1} shows several typical two-qubits gates along with their local variants. Actually, these local invariants are related to the coefficients of the characteristic polynomial of the unitary and symmetric matrix $M(U)$, defined as $M(U)=(Q^\dagger U Q)^T Q^\dagger U Q$, where $Q$ represents the transformation matrix from the logical basis to the magic basis \cite{Hill.97}. In the magic basis, any single-qubit operation is represented by a real orthogonal matrix, ensuring that the spectrum of $M$ remains invariant under local operations. Therefore, if two-qubit gates are equivalent under single-qubit operations, they possess identical sets of local invariants.   Fig.~\ref{Fig4}(a) shows the evolution of the  local invariants as a function of time. At the final evolution time $t=20\,\text{ns}$, $U_{\rm{cz}}$ is achieved. Clearly, it is equivalent to the CNOT gate since they share the same set of local invariants $\{G_1 = 0, G_2 = 0, G_3 = 1\}$. It is of great interest that when $t=6.669\,\text{ns}$ we are able to achieve another type of CZ gate $U_{\rm{cz}}'$ according to the figure
since it shares the same  local invariants with $U_{\rm{cz}}$. On the other hand,  when $t=3.342\,\text{ns}$ where the local invariants are $\{G_1 = 0.5, G_2 = 0, G_3 = 2\}$, an equivalent $\sqrt{\rm{CNOT}}$ gate is obtained. The evolution paths of $\sqrt{\rm{CNOT}}$ and $U_{\rm{cz}}'$ are also shown in Fig.~\ref{Fig2}, which corresponds to $A \rightarrow B$ and $A \rightarrow C$, respectively, where $(\chi_{B}, \xi_{B})=(0.013369\pi, \pi3/2)$ and $(\chi_{C}, \xi_{C})=(0.026674\pi, \pi3/2)$. For both these two operations, since their evolution paths are  always along the longitude of the Bloch sphere, the dynamical phase is not accumulated and thus they also posses the geometric property.


We then conduct simulations to assess the gate-fidelity degradation caused by the low-frequency component of $1/f$ charge noise.   Previous comparisons between theory and experimental data have demonstrated the effectiveness of the quasistatic approximation, especially when the pulse durations are not excessively long \cite{ReedPRL_2016,MartinsPRL_2016}.  In this work, we treat the charge noise as quasistatic during the operation time. This quasistatic noise results in fluctuation on the detuning value $\epsilon$ and further leads to error term $\delta J$. Consequently, we introduce a noisy Hamiltonian by substituting the exchange interaction $J$ as $J + \delta J$.  In our simulation, we average the infidelity over the quasistatic noise $\delta J$ from a normal distribution with a standard deviation $\sigma_J$ and zero mean. In addition, to ensure convergence, we have adopted 500 samples for each $\sigma_J$.  In the simulation, the fidelity is defined as
\begin{equation}
	\begin{aligned}
		F=\frac{\operatorname{Tr}\left(U U^{\dagger}\right)+\left(\left|\operatorname{Tr}\left(U_{\text {ideal }}^{\dagger} U\right)\right|\right)^2}{d(d+1)}.
	\end{aligned}
\end{equation}
Here $U$ is the actual operation with noise, while $U_{\text {ideal }}$ is the ideal target operation, and $d=4$ denotes the dimension of Hilbert space of $U$.  In this work, we take  $h_{0}/2\pi= 2\,\text{MHz}$ ($B_{y,L}^{1}/2\pi=4\,\text{MHz}$), which is a normal amplitude value for the microwave field when using  EDSR.  In experiments, to suppress charge noise, the qubit can be implemented by using barrier control near the sweet spots of the DQDs,  where we have $\delta_{J}=0.00426\ J=0.1597\ h_{0}$ ($\delta_{J}/2\pi\simeq0.319 \,\text{MHz}$) \cite{ZhangPRL_2017}. We consider working in the regime $\Delta E_z+\Delta E^1_{z}= 145.15\ h_{0}$ such that the evolution with respect to $C_{2}\tilde{i}$ at the final time is equivalent to an identity operator in $S_2$.   The resulting infidelities for the implemented geometric  gates, are depicted in Fig.~\ref{Fig5}. It is found that the fidelity for the gates $\sqrt{\rm{CNOT}}$, $U_{\rm{cz}}$ and $U_{\rm{cz}}'$ are $99.99\%$,  $99.9\%$ and  $99.95\%$.  On the other hand, one can see that for the CZ gates, i.e., $U'_{\rm{cz}}$ and $U_{\rm{cz}}$ the infidelity induced by the $B_{y,R}^{1}$ can be neglected safely.

\section{iSWAP and Swap GATE by modulating exchange interaction}
\label{sec4}

\subsection{Effective Hamiltonian}
In the previous section, we have demonstrated how to implement two-qubit gates by modulating magnetic field with square pulse. To make it more friendly to the experiments, it is better to implement gates without square pulse shapes.  On the other hand, one can alternatively modulate the exchange interaction via introducing microwave on the detuning value near the sweet spot of $\epsilon_0$ \cite{Nichol.17,Takeda.20}.  In the absence of magnetic fields along $y$ axis, the Hamiltonian in Eq.~(\ref{eq1}) in the diabatic two-qubit basis $\{|\uparrow \uparrow\rangle,|\uparrow \downarrow\rangle,|\downarrow \uparrow\rangle,|\downarrow \downarrow\rangle\}$ can be expressed as
\begin{equation}
\begin{aligned}
H_{d}&=\begin{pmatrix}
\alpha'& 0 & 0 & 0 \\
0 & -\frac{\beta'}{2}-\frac{J}{2} & \frac{J}{2} & 0 \\
0 & \frac{J}{2} & \frac{\beta'}{2}-\frac{J}{2} & 0 \\
0 & 0 & 0 & -\alpha'
\end{pmatrix} \\
& =\begin{pmatrix}
\alpha' & 0 & 0 & 0 \\
0 & -\frac{\beta'}{2} & 0 & 0 \\
0 & 0 & \frac{\beta'}{2} & 0 \\
0 & 0 & 0 & -\alpha'
\end{pmatrix} + \begin{pmatrix}
0 & 0 & 0 & 0 \\
0 & -\frac{J}{2} & \frac{J}{2} & 0 \\
0 & \frac{J}{2} & -\frac{J}{2} & 0 \\
0 & 0 & 0 & 0
\end{pmatrix} \\
& =H^{\prime}+H_J, \\
\end{aligned}
\end{equation}
where $\alpha'=E_z+E^1_z$, $\beta'=(\Delta E_z+\Delta E^1_z)/2$.   By introducing an oscillating voltage to either gate voltage$V_{L}$ or $V_{R}$, such that $\epsilon(t)=\epsilon_{0}+\epsilon_{1}(t)\cos(\omega' t+\phi')$ with $\epsilon_{1}\ll \epsilon_{0}$. Finally, we have $J(t)=J_0(\epsilon_{0})+J_1(t)\cos(\omega' t+\phi')$, where the Rabi frequency $J_1(t)=\frac{\epsilon_1(t)}{2} J^{\prime}\left(\epsilon_0\right)$ \cite{Nichol.17}.   In addition, we consider operate in the regime $\alpha'\gg J,\,\beta\gg J$. This is reasonable since both  $\alpha'$ and $\beta'$ can be as high as $\sim$GHz  \cite{Nichol.17, VargasPRB_2019}. While the exchange value takes $0\leq J\leq J_{\rm{max}}$.  In experiments, $J_1(t)$ can take as several to 100 MHz \cite{Nichol.17,Sigillito.19,Takeda.20}. Therefore,  $H_J$ can be regared as a  perturbed Hamiltonian by taking a small $J$. Then,  $H_{d}$ can be transformed into another form using the interaction picture, determined by a transoformed operator $U_{t}=\exp \left[-i H^{\prime} t\right]$.  When the oscillating frequency $\omega'$ maches $\beta'$, i.e., on resonance,  the Hamiltonian (with RWA effect)  becomes
 \begin{equation}
     H_I=U_{t}^{\dagger} H_J U_{t}=\begin{pmatrix}
0 & 0 & 0 & 0 \\
0 & -\frac{J_0}{2} & \frac{J_1}{4} e^{i \phi'} & 0 \\
0 & \frac{J_1}{4} e^{-i \phi'} & -\frac{J_0}{2} & 0 \\
0 & 0 & 0 & 0
\end{pmatrix},
 \end{equation} Further, we can rewrite $H_I$ in a reduced subspace $S_3$ spanned by  $\{|\uparrow \downarrow\rangle,|\downarrow \uparrow\rangle\}$ as
 \begin{equation}
H_I'=\begin{pmatrix}
0 &\frac{\Omega}{2} e^{i \phi'}  \\
\frac{\Omega}{2} e^{-i \phi'}&0  \\
\end{pmatrix},
\label{H_I'}
 \end{equation}
where $\Omega=J_1/2$.   It is noted that the evolutions corresponding to $H_I'$ and $H_I$ are equivalent up to a local phase $\eta$ in $S_3$ determined by  $\eta=-\int \frac{J_0}{2} dt$.  The total evolution operator  in the lab frame is  $U_{\rm{lab}}=U_{t} U_I$, where $U_I$ is the unitary operator in the interaction picture. Therefore, by implementing a NOT gate using  $H_I'$ and adjusting the values of $U_{t}$ and $\eta$, either iSWAP or SWAP gate can then be obtained.

\subsection{SWAP and iSWAP geometric gate}

\begin{figure}
	\centering
	\includegraphics[width=1\columnwidth]{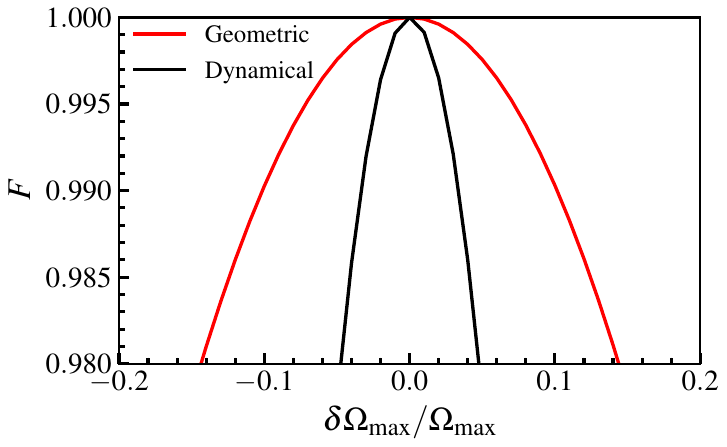}
	\caption{Gate performance of the implemented geometric iSWAP gate compared with the dynamical counterpart realized without using pure geometric phase.}
	\label{Fig6}
\end{figure}

To achieve the geometric NOT gate in subspace $S_3$, one needs to design rotation around the $x$ axis.  To this end, we pick up another two orthogonal time-dependent states as 
{\footnotesize
\begin{equation}
	\begin{aligned}
		|\Phi(t)\rangle & =\mathrm{e}^{-i \frac{f'(t)}{2}}\left[\cos \frac{\chi'}{2} \mathrm{e}^{-i \frac{\xi'(t)}{2}}|\uparrow \downarrow\rangle+\sin \frac{\chi'(t)}{2} \mathrm{e}^{i \frac{\xi'(t)}{2}}|\downarrow \uparrow\rangle\right] \\
		\left|\Phi_{\perp}(t)\right\rangle & =\mathrm{e}^{i \frac{f'(t)}{2}}\left[-\sin \frac{\chi'(t)}{2} \mathrm{e}^{-i \frac{\xi'(t)}{2}}|\uparrow \downarrow\rangle+\cos \frac{\chi'(t)}{2} \mathrm{e}^{i \frac{\xi'(t)}{2}}|\downarrow \uparrow\rangle\right]
	\end{aligned}
\end{equation}
}and divide the evolution loop for the two parameters $\chi'$ and $\xi'$ into four segments \cite{XuFP_2020}
\begin{equation}
	\begin{aligned}
		t \in[0, \tau / 4]: \chi'_1(t) & =\pi\left[1+\sin ^2(2 \pi t / \tau)\right] / 2, \\
		\xi'_1(0) & =0 \\
		t \in[\tau / 4, \tau / 2]: \chi'_2(t) & =\pi\left[1+\sin ^2(2 \pi t / \tau)\right] / 2, \\
		\xi'_2(\tau / 4) & =\xi'_1(\tau / 4)-\gamma, \\
		t \in[\tau / 2,3 \tau / 4]: \chi'_3(t) & =\pi\left[1-\sin ^2(2 \pi t / \tau)\right] / 2, \\
		\xi'_3(\tau / 2) & =\xi'_2(\tau / 2), \\
		t \in[3 \tau / 4, \tau]: \chi'_4(t) & =\pi\left[1-\sin ^2(2 \pi t / \tau)\right] / 2, \\
		\xi'_4(3 \tau / 4) & =\xi'_3(3 \tau / 4)+\gamma,
	\end{aligned}
\label{fequas}
\end{equation}
where $\xi'_j(t)=-\int \dot{f'_j}(t)\cos[\chi'_j(t)]dt$ and $f'_j(t)=\cos[2\chi'_j(t)]/5$ for $j$th part.  The evolution of the state $|\Phi(t)\rangle$ corresponding to Eq.~(\ref{fequas}) on the Bloch sphere is visualized  in Fig.~\ref{Fig2}(b), where it accomplishes cyclic evolution along the path $A \rightarrow B \rightarrow C \rightarrow D \rightarrow A$.  Different from the case in section.~\ref{GQC}, the state $|\Phi(t)\rangle$ does not evolve always along the longitude line. Nevertheless, the accumulated total dynamical phase is still zero.   At the end of the evolution, $|\Phi(t)\rangle$ ($\left|\Phi_{\perp}(t)\right\rangle$) obtains a global geometric phase $\gamma'$ (-$\gamma'$). Then, we have a  geometric operator in subspace $S_3$ as
	\begin{equation}
		\begin{aligned}
			U'(\tau)&=e^{i \gamma'}\left|\Phi_{+}(0)\right\rangle\left\langle\Phi_{+}(0)|+e^{-i \gamma}| \Phi_{\perp}(0)\right\rangle\left\langle\Phi_{\perp}(0)\right| \\
			& =e^{i \gamma \sigma_{x}}.
		\end{aligned}
	\end{equation}
 By setting  $\gamma' = \pi/2$ we can  construct a NOT gate in $S_3$.  In addition, when we take $\alpha', \frac{\beta'}{2}=2n\pi$ with $n$ being an integer number, $U_t$ is thus an identity. In this way, by taking $\eta=2n\pi$ an iSWAP gate is obtained as
 \begin{equation}
	\begin{aligned}
		U_{\rm{iSWAP}}&=\begin{pmatrix}
			1 & 0 & 0 & 0 \\
			0 & 0 & i & 0 \\
			0 & i& 0 &0 \\
			0 & 0 & 0 & 1
		\end{pmatrix}.
	\end{aligned}
\label{iswap}
\end{equation} If we take $\eta=3\pi/2+2n\pi$, a SWAP gate is implemented as
 \begin{equation}
	\begin{aligned}
		U_{\rm{SWAP}}&=\begin{pmatrix}
			1 & 0 & 0 & 0 \\
			0 & 0 & 1& 0 \\
			0 & 1& 0 &0 \\
			0 & 0 & 0 & 1
		\end{pmatrix}.
	\end{aligned}
	\label{swap}
\end{equation} On the other hand, the driving parameters $\Omega(t), \phi'(t)$ related to Eq.~(\ref{fequas}) are inversely obtained as 
 \begin{equation}
 	\label{}
 	\begin{aligned}
 		\Omega(t) & =-\frac{\dot{\chi'}(t)}{\sin (\xi'(t)+\phi'(t))}, \\
 		\phi'(t) & =\arctan \left(\frac{\dot{\chi'}(t)}{\dot{\xi'}(t)} \cot \chi'(t)\right)-\xi'(t),
 	\end{aligned}
 \end{equation} where the pulse shapes of the two parameters are shown in Fig.~\ref{Fig3}(b), where we can see that the pulse shape of $\Omega(t)$ is similar as sine form.

Then, we perform numerical simulation to assess the gate performance. In the simulation,  the maximum value of $\Omega(t)$ is set to be $\Omega_{\text{max}}=2\pi\times 50\,\text{MHz}$.  In Fig.~\ref{Fig4}(b), we plot the local invariant for the gate. At the final gate time ($\tau=32.7\,\text{ns}$)  we find the local invariants are $\{G_1 = 0, G_2 = 0, G_3 = -1\}$. No doubt that an iSWAP gate is realized.  According to the local invariants values,  there is no other entangling gate during the evolution.  The performance of the geometric gate compared with its dynamical counterpart, which is implemented directly using Eq.~(\ref{H_I'}) is shown in Fig.~\ref{Fig6}. To compare fairly we take the same pulse shape of $\Omega(t)$ (the sign might be different) for the two types of gates. It is clear that the robustness of geometric gate outperforms the dynamical one significantly.

\section{Conclusions}
\label{sec5}
In this work, we have proposed feasible way to implement geometric two-qubit gates for silicon-based spin qubits. We consider working in the region where  the exchange interaction is much larger than the local magnetic field gradient. By designing microwave control on either the magnetic field or the exchange interaction, we are able to implement  entangling and non-entangling two-qubit gates in the considered subspace.  Employing operations with pure geometric phase, this work predicts two-qubit gates with fidelity over 99\% under the noise level related to the experiments. In addition,  the superiority of the geometric gates are varified comparing to the dynamical gate, showcasing their significant potential  application for robust quantum computing.

	\begin{acknowledgments}
	This work was supported by the National Natural Science
	Foundation of China (Grant No. 11905065, 62171144, 12305019), and
	the Guangxi Science Foundation (Grant No. AD22035186,
	2021GXNSFAA220011).
\end{acknowledgments}

\appendix

\setcounter{equation}{0}

\section{Aharonov and Anandan phase}
\label{append}

Aharonov and Anandan \cite{AharonovPRL_1987} demonstrated that any cyclic evolution beginning from state $\ket{\psi(0)}$ and concluding in the parallel state $\ket{\psi(T)}$ at time $T$ is associated with a geometric phase. This phase is defined as follows:
\begin{equation}
	\label{geo}
	\begin{aligned}
		\Gamma_g(T) & =\arg \bra{\zeta(0)}\zeta(T)\rangle+i \int_0^T\bra{\zeta(t)} \dot{\zeta}(t)\rangle d t \\
		& =i \int_0^T\left\langle \Theta(t)\left|\partial_t\right| \Theta(t)\right\rangle d t
	\end{aligned}
\end{equation}
The first line of Eq.~(\ref{geo}) involves subtracting the dynamical phase $\Gamma_d$ from the total phase $\Gamma_{tot}$, denoted as $\Gamma _g(T) = \Gamma_{tot}(T) - \Gamma_d(T)$. The dynamical phase $\Gamma_d$ is commonly expressed as an integral of the expectation value of the Hamiltonian, $\bra{\zeta(t)}H(t)\ket{\zeta(t)}$ which is frame-dependent \cite{DongPRXQuantum_2021}. 

The second line of Eq.(\ref{geo}) arises from transforming $\ket{\zeta(t)}$ into the state $\ket{\Theta(t)}$ within a projective space. This space is defined in such a way that all parallel states $\ket{\zeta(t)}$ map to the same $\ket{\Theta(t)}$. In the context of this study, focusing on a two-level system, $\ket{\Theta(t)}$ resides on the Bloch sphere. The geometric phase is determined by the area enclosed within the trajectory of $\ket{\Theta(t)}$. It's important to highlight that, unlike the Berry phase \cite{Berry_1984}, the Aharonov-Anandan phase doesn't require an adiabatic condition; it holds exact for any cyclic evolution. From this point onward, we will refer to the Aharonov-Anandan phase as the geometric phase.

The distinctive characteristic of the geometric phase is its dependence solely on the path followed by $\ket{\Theta(t)}$ on the Bloch sphere \cite{SamuelPRL_1988,Pachos_2001}.
Consequently, the geometric phase remains unaffected by "parallel" errors,
which only impact the rate at which the path is traversed without altering its shape \cite{ChiaraPRL_2003,ZhuPRA_2005}.
As detailed in Ref.\cite{DongPRXQuantum_2021}, this property renders the geometric phase resilient against first- and second-order parallel errors, including specific errors arising from driving field issues. It's worth noting that, in contrast, the dynamical phase is generally vulnerable to these errors, even at the first order. The resilience of geometric phases against errors forms the core concept of holonomic quantum computation \cite{ZanardiPLA_1999}.


%

\end{document}